\documentstyle[12pt]{article}
\def\be{\begin{equation}}
\def\ee{\end{equation}}
\def\l{\label}

\def\S{{\cal S}}

\def\W{{\cal W}}

\makeatletter \ifcase\@ptsize \font\teneufm=eufm10
\font\seveneufm=eufm7 \font\fiveeufm=eufm5
\font\teneusm=eusm10 \font\seveneusm=eusm7
\font\fiveeusm=eusm5 \or \font\teneufm=eufm10 scaled
\magstephalf \font\seveneufm=eufm7 \font\fiveeufm=eufm5
\font\teneusm=eusm10 scaled \magstephalf
\font\seveneusm=eusm7 \font\fiveeusm=eusm5 \or
\font\teneufm=eufm10 scaled \magstep1 \font\seveneufm=eufm7
\font\fiveeufm=eufm5 \font\teneusm=eusm10 scaled \magstep1
\font\seveneusm=eusm7 \font\fiveeusm=eusm5 \fi

\newfam\eufmfam \newfam\eusmfam \textfont\eufmfam=\teneufm
\scriptfont\eufmfam=\seveneufm
\scriptscriptfont\eufmfam=\fiveeufm
\textfont\eusmfam=\teneusm \scriptfont\eusmfam=\seveneusm
\scriptscriptfont\eusmfam=\fiveeusm

\def\frak{\ifmmode\let\next\frak@\else
 \def\next{\errmessage{Use \string\frak\space only in math
 mode}}\fi\next} \def\frak@#1{{\frak@@{#1}}}
 \def\frak@@#1{\fam\eufmfam#1} 
 \def\sh{\ifmmode\let\next\sh@\else
 \def\next{\errmessage{Use \string\sh\space only in math
 mode}}\fi\next} \def\sh@#1{{\sh@@{#1}}}
 \def\sh@@#1{\fam\eusmfam#1}

\ifcase\@ptsize \font\tenmsa=msam10 \font\sevenmsa=msam7
 \font\fivemsa=msam5 \font\tenmsb=msbm10
 \font\sevenmsb=msbm7 \font\fivemsb=msbm5 \or
 \font\tenmsa=msam10 scaled \magstephalf
 \font\sevenmsa=msam7 \font\fivemsa=msam5
 \font\tenmsb=msbm10 scaled \magstephalf
 \font\sevenmsb=msbm7 \font\fivemsb=msbm5 \or
 \font\tenmsa=msam10 scaled \magstep1 \font\sevenmsa=msam7
 \font\fivemsa=msam5 \font\tenmsb=msbm10 scaled \magstep1
 \font\sevenmsb=msbm7 \font\fivemsb=msbm5 \fi

\newfam\msafam \newfam\msbfam \textfont\msafam=\tenmsa
\scriptfont\msafam=\sevenmsa
\scriptscriptfont\msafam=\fivemsa \textfont\msbfam=\tenmsb
\scriptfont\msbfam=\sevenmsb
\scriptscriptfont\msbfam=\fivemsb

\def\Bbb{\ifmmode\let\next\Bbb@\else
 \def\next{\errmessage{Use \string\Bbb\space only in math
 mode}}\fi\next} \def\Bbb@#1{{\Bbb@@{#1}}}
 \def\Bbb@@#1{\fam\msbfam#1} \def\hexnumber@#1{\ifnum#1<10
 \number#1\else \ifnum#1=10 A\else\ifnum#1=11
 B\else\ifnum#1=12 C\else \ifnum#1=13 D\else\ifnum#1=14
 E\else\ifnum#1=15 F\fi\fi\fi\fi\fi\fi\fi}
 \def\msa@{\hexnumber@\msafam} \def\msb@{\hexnumber@\msbfam}
 \mathchardef\square="0\msa@03

\makeatother
 \newcommand{\RR}{{\Bbb R}}

\begin{document}
\begin{titlepage}

\rightline{UMN-TH-1722-98-TPI-MINN-98/19}
\rightline{DFPD97/TH/51}
\rightline{September 1998, \tt hep-th/9809125}

\vspace{0.333cm}

\begin{center}

{\Large \bf Equivalence Principle, Planck Length and}

\vspace{.333cm}

{\Large \bf Quantum Hamilton--Jacobi Equation}

\vspace{.9333cm}

{\large Alon E. Faraggi$^{1}$ $\,$and$\,$ Marco Matone$^{2}$\\}
\vspace{.2in}
{\it $^{1}$ Department of Physics \\
University of Minnesota, Minneapolis MN 55455, USA\\
e-mail: faraggi@mnhepo.hep.umn.edu\\}
\vspace{.02in}
{\it $^{2}$ Department of Physics ``G. Galilei'' -- Istituto
Nazionale di Fisica Nucleare\\
University of Padova, Via Marzolo, 8 -- 35131 Padova, Italy\\
e-mail: matone@padova.infn.it\\}

\end{center}

\vspace{.333cm}

\centerline{\large \bf Abstract}

\vspace{.333cm}

The Quantum Stationary HJ Equation (QSHJE) that we derived from the equivalence
principle, gives rise to initial conditions which cannot be seen in the
Schr\"odinger equation. Existence of the classical limit leads to a dependence
of the integration constant $\ell=\ell_1+i\ell_2$ on the Planck length.
Solutions of the QSHJE provide a trajectory representation of quantum mechanics
which, unlike Bohm's theory, has a non--trivial action even for bound states and
no wave guide is present. The quantum potential turns out to be an intrinsic
potential energy of the particle which, similarly to the relativistic rest
energy, is never vanishing.

\vspace{3.33cm}



\end{titlepage}
\setcounter{footnote}{0}
\renewcommand{\thefootnote}{\arabic{footnote}}

\newpage

Let us consider a one--dimensional stationary system of energy $E$ and
potential $V$ and set $\W\equiv V(q)-E$. In \cite{1} the following equivalence
principle has been formulated

\vspace{.333cm}

\noindent
{\it For each pair $\W^a,\W^b$, there is a transformation
$q^a\longrightarrow q^b=v(q^a)$, such that}
\be
\W^a(q^a)\longrightarrow {\W^a}^v (q^b)=\W^b(q^b).
\l{equivalence}\ee

\vspace{.333cm}

Implementation of this principle uniquely leads to the Quantum
Stationary HJ Equation (QSHJE) \cite{1}
\be
{1\over 2m}\left({\partial\S_0(q)\over \partial q}\right)^2+V(q)-E
+{\hbar^2\over 4m}\{\S_0,q\}=0,
\l{1Q}\ee
where $\S_0$ is the Hamilton's characteristic function also called reduced
action. In this equation the Planck constant plays the role of covariantizing
parameter. The fact that a fundamental constant follows from the equivalence
principle suggests that other fundamental constants as well may be related to
such a principle.

We have seen in \cite{1} that the implementation of the equivalence principle
implied a cocycle condition which in turn determines the structure of the
quantum potential. A property of the formulation is that unlike in Bohm's theory
\cite{Bohm}\cite{Holland}, the quantum potential
\be
Q(q)={\hbar^2\over 4m}\{\S_0,q\},
\l{quqq}\ee
like $\S_0$, is never trivial. This reflects in the fact that a general solution
of the Schr\"odinger equation will have the form
\be
\psi={1\over\sqrt{\S_0'}}\left(Ae^{-{i\over\hbar}\S_0}+Be^{{i\over\hbar}\S_0}
\right).
\l{opiYh9}\ee
If $(\psi^D,\psi)$ is a pair of real linearly independent solutions
of the Schr\"odinger equation, then we have
\be
e^{{2i\over \hbar}\S_0}=e^{i\alpha}{w+i\bar\ell\over w-i\ell},
\l{solls0}\ee
where $w=\psi^D/\psi$, and ${\rm Re}\,\ell\ne 0$. We note that $\ell$, on which
the dynamics depends, does not appear in the conventional formulation of quantum
mechanics.

In this Letter we show that non--triviality of the quantum potential of the free
particle with vanishing energy is at the heart of the existence of a length
scale. Actually, we will show the appearance of the Planck length in the
complex integration constant $\ell$, indicating that gravity is intrinsically
and deeply connected with quantum mechanics. Therefore, there is trace of
gravity in the constant $\ell$ whose role is that of initial condition for the
dynamical equation (\ref{1Q}). In this context we stress that $\ell$ plays the
role of ``hidden'' constant as it does not appear in the Schr\"odinger equation.
This is a consequence of the fact that whereas the Schr\"odinger equation is a
second--order linear differential equation, the QSHJE is a third--order 
non--linear one, with the associated dynamics being deeply connected to the
M\"obius symmetry of the Schwarzian derivative.

Before going further it is worth stressing that the basic difference between our
$\S_0$ and the one in Bohm's theory \cite{Bohm}\cite{Holland} arises for bound
states. In this case the wave function $\psi$ is proportional to a real
function. This implies that with Bohm's identification $\psi(q)=R(q)\exp(i
{\S}_0/\hbar)$, one would have $\S_0=cnst$. Therefore, all bound states, like
in the case of the harmonic oscillator, would have $\S_0=cnst$ (for which the
Schwarzian derivative is not defined). Besides the difficulties in getting a
non--trivial classical limit for $p=\partial_q\S_0$, this seems an
unsatisfactory feature of Bohm's theory which completely disappears if one uses
Eq.(\ref{opiYh9}). This solution directly follows from the QSHJE: reality of
$\psi$ simply implies that $|A|^2=|B|^2$ and there is no trace of the solution
$\S_0=cnst$ of Bohm theory. Furthermore, we would like to remark that in Bohm's
approach some interpretational aspects are related to the concept of a
pilot--wave guide. There is no need for this in the present formulation. This
aspect and related ones have been investigated also by Floyd \cite{Floyd}.
Nevertheless, there are some similarities between the approach and Bohm's
interpretation of quantum mechanics. In particular, solutions of the QSHJE
provide a trajectory representation of quantum mechanics. That is for a given
quantum mechanical system, by solving the QSHJE for $\S_0(q)$, we can evaluate
$p=\partial_q\S_0(q)$ as a function of the initial conditions. Thus, for a given
set of initial conditions we have a predetermined orbit in phase space. The
solutions to the third--order non--linear QSHJE are obtained by utilizing the
two linearly independent solutions of the corresponding Schr\"odinger equation.

In the case of the free particle with vanishing energy, $i.e.$ with $\W(q)=
\W^0(q^0)\equiv 0$, we have that two real linearly independent solutions of
the Schr\"odinger equation are $\psi^{D^0}=q^0$ and $\psi^0=1$. As the different
dimensional properties of $p$ and $q$ led to introduce the Planck constant
in the QSHJE \cite{1}, in the case of $\psi^{D^0}$ and $\psi^0$ we have to
introduce a length constant.

Let us derive the quantum potential in the case of the state $\W^0$.
By (\ref{quqq}) and (\ref{solls0}) we have
\be
Q^0=-{p_0^2\over 2m}=-{\hbar^2(\ell_0+\bar\ell_0)^2\over 8m|q^0-i\ell_0|^4},
\l{oix9ui87}\ee
where $p_0=\partial_{q^0}\S_0^0(q^0)$. Therefore, we see that the quantum
potential is an intrinsic property of the particle as even in the case of
the free particle of vanishing energy one has $Q_0\not\equiv 0$. This is
strictly related to the local homeomorphicity properties that the ratio of any
real pair of linearly independent solutions of the Schr\"odinger equation should
satisfy, a consequence of the M\"obius symmetry of the
Schwarzian derivative \cite{5}. In this context let us recall that 
$\{\S_0,q\}$ is not defined for $\S_0=cnst$.
More generally, the QSHJE is well defined if and only if the corresponding
$w\ne cnst$ is of class $C^2(\hat\RR)$ with $\partial_q^2w$ differentiable on
the extended real line $\hat\RR=\RR\cup\{\infty\}$ \cite{5}.

We stress that $Q$ does not correspond to the Bohm's quantum potential which, in
a different context, was considered as internal potential in
\cite{MugaSalaSnider}.
Here we have the basic fact that the quantum potential is never trivial, an
aspect which is strictly related to $p$--$q$ duality and to the existence of the
Legendre transformation of $\S_0$ for any $\W$ \cite{1}\cite{5}.

Besides the different dimensionality of $\psi^{D^0}$ and $\psi^0$, we also note
that similarity between the equivalence principle we formulated and the one at
the heart of general relativity would suggest the appearance of some other
fundamental constants besides the Planck constant. We now show that
non--triviality of $Q^0$ or, equivalently, of $p_0$, has an important
consequence in considering the classical and $E\longrightarrow 0$ limits in
the case of the free particle. In doing this, we will see the appearance of
the Planck length in the complex integration constant $\ell$ of the QSHJE.

Let us consider the conjugate momentum in the case of the free particle
of energy $E$. We have \cite{5}
\be
p_E=\pm{\hbar (\ell_E+\bar\ell_E)\over 2|k^{-1}\sin kq-i\ell_E \cos kq|^2},
\l{dajjj}\ee
where $k=\sqrt{2mE}/\hbar$. The first condition is that in the $\hbar
\longrightarrow 0$ limit the conjugate momentum reduces to the classical one
\be
\lim_{\hbar\longrightarrow 0}p_E=\pm\sqrt{2mE}.
\l{bos1S11b}\ee
On the other hand, we should also have
\be
\lim_{E\longrightarrow 0}p_E=p_0=\pm {\hbar
(\ell_0+\bar\ell_0)\over 2|q-i\ell_0|^2}.
\l{bisCS11b}\ee

Let us first consider the limit (\ref{bos1S11b}). By (\ref{dajjj}) we see that
in order to reach the classical value $\sqrt{2mE}$ in the $\hbar
\longrightarrow 0$ limit, the quantity $\ell_E$ should depend on $E$. Let us set
\be
\ell_E= k^{-1}f(E,\hbar)+\lambda_E,
\l{prova4}\ee
where $f$ is dimensionless. Since $\lambda_E$ is still arbitrary, we can choose
$f$ to be real. By (\ref{dajjj}) we have
\be
p_E=\pm{\sqrt{2mE}f(E,\hbar)+mE(\lambda_E+\bar\lambda_E)/\hbar\over
\left|e^{ikq}+(f(E,\hbar)-1+\lambda_E k )\cos kq\right|^2}.
\l{prova4cc}\ee
Observe that if one ignores $\lambda_E$ and sets $\lambda_E=0$, then
by (\ref{bos1S11b}) we have
\be
\lim_{\hbar\longrightarrow 0} f(E,\hbar)=1.
\l{prova5}\ee
We now consider the properties that $\lambda_E$ and $f$ should have in order
that (\ref{prova5}) be satisfied in the physical case in which $\lambda_E$ is
arbitrary but for the condition ${\rm Re}\,\ell_E\ne 0$, as required by the
existence of the QSHJE. First of all note that cancellation of the divergent
term $E^{-1/2}$ in
\be
p_E\; {}_{\stackrel{\sim}{E\longrightarrow 0}} \pm{2\hbar^2(2mE)^{-1/2}f(E,
\hbar)+\hbar(\lambda_E+\bar\lambda_E)\over 2|q -i\hbar(2mE)^{-1/2}f(E,\hbar)-i
\lambda_E|^2},
\l{prova3truciolo}\ee
yields
\be
\lim_{E\longrightarrow 0}E^{-1/2}f(E,\hbar)=0.
\l{prova6}\ee

The limit (\ref{bisCS11b}) can be seen as the limit in which the trivializing
map reduces to the identity. Actually, the trivializing map, introduced in
\cite{1} and further investigated in \cite{5}, which connects the
state $\W=-E$ with the state $\W^0$, reduces to the identity map in the
$E\longrightarrow 0$ limit. In the above investigation we considered $q$ as
independent variable, however one can also consider $q_E(q^0)$ so that $\lim_{E
\to 0} q_E=q^0$ and in the above formulas one can replace $q$ with $q_E$.

We know from (\ref{prova6}) that $k$ must enter in the expression of $f(E,
\hbar)$. Since $f$ is a dimensionless constant, we need at least one more
constant with the dimension of a length. Two fundamental lengths one can
consider are the Compton length
\be
\lambda_c={\hbar \over mc},
\l{prova7}\ee
and the Planck length
\be
\lambda_p=\sqrt{\hbar G\over c^3}.
\l{prova8}\ee
Two dimensionless quantities depending on $E$ are
\be
x_c=k\lambda_c=\sqrt{2E\over mc^2},
\l{kComptlength}\ee
and
\be
x_p=k\lambda_p=\sqrt{2mEG\over\hbar c^3}.
\l{kPlancklength}\ee
On the other hand, concerning $x_c$ we see that it does not depend on $\hbar$
so that it cannot be used to satify (\ref{prova5}). Therefore, we see that a
natural expression for $f$ is a
function of the Planck length times $k$. Let us set
\be
f(E,\hbar)=e^{-\alpha(x_p^{-1})},
\l{prova9}\ee
where
\be
\alpha(x_p^{-1}) =\sum_{k\geq 1} \alpha_kx_p^{-k}.
\l{alphapp}\ee
The conditions (\ref{prova5})(\ref{prova6}) correspond to conditions on the
coefficients $\alpha_k$. For example, in the case in which one considers
$\alpha$ to be the function 
\be
\alpha(x_p^{-1}) =\alpha_1x_p^{-1},
\l{cdtnn1}\ee
then by (\ref{prova6}) we have 
\be
\alpha_1>0.
\l{cdtneee}\ee

In order to consider the structure of $\lambda_E$, we note that although
$e^{-\alpha(x_p^{-1})}$ cancelled the $E^{-1/2}$ divergent term, we still have
some conditions to be satisfied. To see this note that
\be
p_E=\pm{\sqrt{2mE}e^{-\alpha(x_p^{-1})}+mE(\lambda_E+\bar\lambda_E)/\hbar
\over\left|e^{ikq}+(e^{-\alpha(x_p^{-1})}-1+k\lambda_E)\cos kq\right|^2},
\l{prova20}\ee
so that the condition (\ref{bos1S11b}) implies
\be
\lim_{\hbar\longrightarrow 0}{\lambda_E\over\hbar}=0.
\l{prova21}\ee
To discuss this limit, we first note that
\be
p_E=\pm{2\hbar k^{-1}e^{-\alpha(x_p^{-1})}+\hbar(\lambda_E+\bar\lambda_E)\over
2\left|k^{-1}\sin kq-i\left(k^{-1}e^{-\alpha(x_p^{-1})}+\lambda_E\right)\cos
kq\right|^2}.
\l{prova21bbv}\ee
So that, since
\be
\lim_{E\longrightarrow 0}k^{-1}e^{-\alpha(x_p^{-1})}=0,
\l{prova21ccv}\ee
we have by (\ref{bisCS11b}) and (\ref{prova21bbv}) that
\be
\lambda_0=
\lim_{E\longrightarrow 0}{\lambda_E}=\lim_{E\longrightarrow 0}{\ell_E}=\ell_0.
\l{prova22}\ee

Let us now consider the limit
\be
\lim_{\hbar\longrightarrow 0} p_0=0.
\l{prova23}\ee
First of all note that, since
\be
p_0=\pm {\hbar(\ell_0+\bar\ell_0)\over 2|q^0-i\ell_0|^2},
\l{prova24}\ee
we have that the effect on $p_0$ of a shift of ${\rm Im}\, \ell_0$
is equivalent to a shift of the coordinate. Therefore, in considering the
limit (\ref{prova23}) we can set  ${\rm Im}\, \ell_0=0$ and distinguish
the cases $q^0\ne 0$ and $q^0=0$. Observe that as we always have
${\rm Re}\,\ell_0\ne 0$,
it follows that the denominator in the right hand side of (\ref{prova24})
is never vanishing. Let us define $\gamma$ by
\be
{\rm Re}\,\ell_0
{}_{\stackrel{\sim}{\hbar\longrightarrow 0}}\hbar^{\gamma}.
\l{ioq990p}\ee
We have
\be
p_0 {}_{\stackrel{\sim}{\hbar\longrightarrow 0}}
\left\{\begin{array}{ll} \hbar^{\gamma+1}, & q_0\ne 0,\\ \hbar^{1-\gamma},
& q_0=0, \end{array}\right.
\l{Vu1fy9}\ee
and by (\ref{prova23})
\be
-1<\gamma<1.
\l{prova25}\ee
A constant length having powers of $\hbar$ can be constructed by means of 
$\lambda_c$ and $\lambda_p$. We also note that a constant length
independent from $\hbar$ is provided by $\lambda_e=e^2/mc^2$ where $e$ is the
electric charge. Then $\ell_0$ can be considered as a suitable function of
$\lambda_c$, $\lambda_p$ and $\lambda_e$ satisfying the constraint
(\ref{prova25}).

The above investigation indicates that a natural way to express $\lambda_E$ is
given by
\be
\lambda_E=e^{-\beta(x_p)}\lambda_0,
\l{prova26}\ee
where 
\be
\beta(x_p)=\sum_{k\geq 1}\beta_k x_p^k.
\l{prova267Y}\ee
Any possible choice of $\beta(x_p)$ should satisfy the conditions
(\ref{prova21}) and (\ref{prova22}). For example, for the modulus $\ell_E$
built with $\beta(x_p)=\beta_1 x_p$, one should have $\beta_1>0$.

Summarizing, by (\ref{prova4})(\ref{prova9})(\ref{prova22}) and (\ref{prova26})
we have
\be
\ell_E= k^{-1}e^{-\alpha(x_p^{-1})}+e^{-\beta(x_p)}\ell_0,
\l{prova27}\ee
where
$\ell_0=\ell_0(\lambda_c,\lambda_p,\lambda_e)$, and for the conjugate
momentum of the state $\W=-E$ we have
\be
p_E=\pm{2k^{-1}\hbar e^{-\alpha(x_p^{-1})}+\hbar e^{-\beta(x_p)}(\ell_0+\bar
\ell_0)\over 2\left|k^{-1}\sin kq-i\left(k^{-1}e^{-\alpha(x_p^{-1})}+e^{-\beta
(x_p)}\ell_0\right)\cos kq\right|^2}.
\l{prova21bbvxx}\ee

It is worth recalling that the conjugate momentum does not correspond to the
mechanical momentum $m\dot q$ \cite{Floyd}\cite{5}. However, the velocity
itself is strictly related to it. In particular, following the suggestion by
Floyd \cite{Floyd} of using Jacobi's theorem to define time parametrization,
we have that velocity and conjugate momentum satisfy the relation \cite{5}
\be
\dot q={1\over\partial_E p},
\l{doiqI}\ee
which holds also classically.

We stress that the appearance of the Planck length is strictly related to
$p$--$q$ duality and to the existence of the Legendre transformation of $\S_0$
for any state. This $p$--$q$ duality has a counterpart in the $\psi^D$--$\psi$
duality \cite{1}\cite{5} which sets a length scale which already appears in
considering linear combinations of $\psi^{D^0}=q^0$ and $\psi^0=1$. This aspect 
is related to the fact that we always have $\S_0\ne cnst$ and $\S_0\not\propto
q+cnst$, so that also for the states $\W^0$ and $\W=-E$ one has a non--constant
conjugate momentum. In particular, the Planck length naturally emerges in
considering $\lim_{E\to 0}p_E=p_0$, together with the analysis of the
$\hbar\longrightarrow 0$ limit of both $p_E$ and $p_0$.

\vspace{.333cm}

Work supported in part by DOE Grant No.\ DE--FG--0287ER40328 (AEF)
and by the European Commission TMR programme ERBFMRX--CT96--0045 (MM).

\end{document}